\documentclass[sigplan,screen,nonacm]{acmart}
\AtBeginDocument{%
  }

\usepackage[most]{tcolorbox} 

\newtcolorbox{finding}[1]{
  sharp corners,
  enhanced,
  boxrule=0pt,
  left=10pt,
  right=5pt,
  top=5pt,
  bottom=5pt,
  colback=gray!5, 
  borderline west={2pt}{0pt}{gray!80}, 
  title=\textbf{#1},
  coltitle=black,
  detach title,
  before upper={\textcolor{black}{\textbf{#1}} \itshape}, 
  fontupper=\small,
  boxsep=0pt,
  before skip=10pt,
  after skip=10pt
}

\usepackage{booktabs}
\usepackage{multirow}
\usepackage{array}
\usepackage{subfigure}
\usepackage{graphicx}

\tcbuselibrary{listings, breakable} 
\setcopyright{acmlicensed}
\copyrightyear{2026}
\acmYear{2026}
\acmDOI{XXXXXXX.XXXXXXX}
\acmConference[Arxiv]{Make sure to enter the correct
  conference title from your rights confirmation email}{}{preprint}

\begin{document}

\title[]{From Craft to Kernel: A Governance-First Execution Architecture and Semantic ISA for Agentic Computers}

\author{Xiangyu Wen}
\affiliation{%
  \institution{CUHK} 
  \country{Hong Kong SAR, China}
}

\author{Yuang Zhao}
\affiliation{%
  \institution{Shanghai Jiao Tong University}
  \country{Shanghai, China}
}

\author{Xiaoyu Xu}
\affiliation{%
  \institution{Zhejiang University}
  \country{Hangzhou, China}
}

\author{Lingjun Chen}
\affiliation{%
 \institution{Peking University}
 \country{Beijing, China}
}

\author{Changran Xu}
\affiliation{%
  \institution{CUHK}
  \country{Hong Kong SAR, China}
}

\author{Shu Chi}
\affiliation{%
  \institution{Tsinghua University}
  \country{Beijing, China}
}

\author{Jianrong Ding}
\affiliation{%
  \institution{CUHK}
  \country{Hong Kong SAR, China}
}

\author{Zeju Li}
\affiliation{%
  \institution{CUHK}
  \country{Hong Kong SAR, China}
}

\author{Haomin Li}
\affiliation{%
  \institution{Shanghai Jiao Tong University}
  \country{Shanghai, China}
}

\author{Li Jiang}
\affiliation{%
  \institution{Shanghai Jiao Tong University}
  \country{Shanghai, China}
}

\author{Fangxin Liu}
\affiliation{%
  \institution{Shanghai Jiao Tong University}
  \country{Shanghai, China}
}

\author{Qiang Xu}
\authornote{Corresponding author. Email: qxu@cse.cuhk.edu.hk.}
\affiliation{%
  \institution{CUHK}
  \country{Hong Kong SAR, China}
}

\renewcommand{\shortauthors}{Wen et al.}

\begin{abstract}
  The transition of agentic AI from brittle prototypes to production systems is stalled by a pervasive crisis of craft. We suggest that the prevailing orchestration paradigm—delegating the system control loop to large language models and merely patching with heuristic guardrails—is the root cause of this fragility. Instead, we propose Arbiter-K, a Governance-First execution architecture that reconceptualizes the underlying model as a Probabilistic Processing Unit encapsulated by a deterministic, neuro-symbolic kernel. Arbiter-K implements a Semantic Instruction Set Architecture (ISA) to reify probabilistic messages into discrete instructions. This allows the kernel to maintain a Security Context Registry and construct an Instruction Dependency Graph at runtime, enabling active taint propagation based on the data-flow pedigree of each reasoning node. By leveraging this mechanism, Arbiter-K precisely interdicts unsafe trajectories at deterministic sinks (e.g., high-risk tool calls or unauthorized network egress) and enables autonomous execution correction and architectural rollback when security policies are triggered. Evaluations on OpenClaw and NanoBot demonstrate that Arbiter-K enforces security as a microarchitectural property, achieving 76\% to 95\% unsafe interception for a 92.79\% absolute gain over native policies. The code is publicly available at \url{https://github.com/cure-lab/ArbiterOS}.
\end{abstract}

\maketitle

\section{Introduction}
The emergence of agentic AI as a primary computational workload has initiated a critical transition from brittle prototypes to production grade systems. Unlike traditional inference tasks that are transient and stateless, agentic workloads involve long running execution traces with deep state dependencies and frequent interactions with the host environment. These systems solve complex objectives through iterative reasoning and acting loops, necessitating the invocation of external tools and the access to sensitive system resources. Given the high degree of autonomy inherent in these agents, their execution trajectories are intrinsically non-deterministic and exert direct side effects on the underlying system state. Consequently, establishing a secure and verifiable execution model for such workloads has emerged as a fundamental reliability challenge.

We suggest that the prevailing Orchestration paradigm is the root cause of this fragility. This design pattern commits a fundamental category error by treating the Large Language Model (LLM) as the core of the system control loop. Conceptually, current frameworks grant an opaque and stochastic inference engine the authority typically reserved for a secure system kernel. This grants the untrusted model authority over critical control flows, making the agent intrinsically vulnerable to cascading errors and semantic injections. Existing security measures function as reactive filters on top of black boxes, offering only local output sanitization without formal guarantees for global state transitions or architectural integrity.

This architectural vulnerability leads to a pervasive crisis of craft where reliability is treated as an emergent property of model behavior rather than a guaranteed result of principled design. This flaw results in state of the art models achieving success rates as low as 30\% on complex real world tasks~\citep{TheAgentCompany2024}. Existing guardrails prove extremely fragile against semantic injection attacks; experimental data indicates that over 40\% of malicious instructions can bypass text based defense mechanisms~\citep{Liu2026ClawKeeperCS}. Furthermore, relying on the LLM itself for security self-checks incurs prohibitive computational overhead. Importantly, by treating agent traces as opaque plain text, existing systems cannot perform fine-grained privilege verification or data-flow auditing.

Our key observation is that the failure of agentic governance is rooted in the absence of a formal interface between probabilistic reasoning and deterministic execution. In classical systems, the Instruction Set Architecture (ISA) serves as the fundamental contract between intent and execution, abstracting operations into a set of discrete primitives with deterministic side effects. We suggest that agentic computing requires a similar contract to bridge this semantic gap. The significance of a Semantic ISA lies in providing the physical basis for instruction decoding and architectural auditing. By reifying opaque token streams into atomic semantic instructions, the kernel can define explicit execution privileges and data dependencies. This transformation allows the system to convert unobservable semantic deviations into captureable architectural exceptions.

To implement this insight, we propose \textbf{Arbiter-K}, a governance first execution architecture. We reconceptualize the LLM as a Probabilistic Processing Unit (PPU) and define a Semantic ISA comprising five specialized logical cores. The architecture demotes the PPU to a non-privileged proposal generator while mandating that all environment altering instructions be validated by a symbolic kernel. The kernel maintains a Security Context Registry and dynamically constructs an Instruction Dependency Graph (IDG) during runtime. This enables an active taint propagation mechanism based on data flow pedigree, ensuring that the kernel can proactively interdict unsafe trajectories before they reach deterministic sinks. Furthermore, \textbf{Arbiter-K} enables the autonomous execution correction, allowing the system to fully reuse the feedback from the security kernel upon detecting semantic divergence.

We implemented a prototype of \textbf{Arbiter-K} on top of the OpenClaw~\citep{openclaw} and NanoBot~\citep{nanobot} frameworks. Our evaluation shows that, by enforcing security as an architectural property, \textbf{Arbiter-K} effectively blocks sophisticated semantic attacks while requiring only minimally invasive changes to the underlying agent frameworks. Preliminary results indicate that \textbf{Arbiter-K} intercepts more than 92\% of unauthorized access attempts, while incurring a false interception rate of less than 6\% on benign operations in the NanoBot framework. These results demonstrate that \textbf{Arbiter-K} can substantially strengthen execution determinism and enforce resource boundaries without degrading the agent’s reasoning capabilities.
Contributions are summarized as follows:
\vspace{-10pt}
\begin{itemize}
    \item We define a Semantic ISA that provides a formal execution abstraction for agentic workloads, bridging the gap between probabilistic reasoning and deterministic system sinks.

    \item We propose \textbf{Arbiter-K}, a governance-first architecture that utilizes an Instruction Dependency Graph and a Security Context Registry to implement active taint propagation and architectural rollback.

    \item We demonstrate through an implementation on the OpenClaw and NanoBot frameworks that Arbiter-K successfully enforces microarchitectural security invariants while maintaining high system performance.
\end{itemize}

\vspace{-10pt}
\section{Backgrounds}
\subsection{Prompt Engineering for Agent Building}
Prompt engineering has transitioned from a superficial interface into the primary mechanism for directing the behavior of PPUs. Current methodologies emphasize structured and multi-stage control flow to maximize performance in complex planning and tool-use tasks~\citep{Dhrif2025ReasoningAwarePO}. This evolution includes techniques such as chain-of-thought and ReAct-style reasoning~\citep{Xu2026AIAS}, which are increasingly treated as a stochastic optimization problem where instructions and exemplars are tuned to steer model output distributions~\citep{Sahoo2024ASS, Xu2026AIAS}.

In sophisticated environments~\citep{Junprung2023ExploringTI}, advanced frameworks for multi-agent coordination demonstrate how higher-level manager agents can synthesize and refine subordinate system prompts by utilizing chain prompting, personas, and self-reflection to scale specialization~\citep{Bo2024ReflectiveMC, Schulhoff2024ThePR, Xu2026AIAS}. Consequently, prompt engineering serves as the central architectural lever for defining agent behavior rather than a superficial interface layer.
\begin{table*}[ht]
\setlength{\abovecaptionskip}{1pt}
\setlength{\belowcaptionskip}{1pt}
\centering
\caption{Interception success rate ($\uparrow$) of native host guardrails under indirect semantic-injection evaluation.}
\label{tab:bypass-rates}
\resizebox{1\linewidth}{!}{
\begin{tabular}{@{}cccccccc@{}}
\toprule
\multicolumn{2}{c}{\textbf{Agent System}}                        & \multicolumn{3}{c}{\textbf{OpenClaw}}                                     & \multicolumn{3}{c}{\textbf{Nanobot}}                                      \\ \midrule
\multicolumn{2}{c}{\textbf{Models}}                              & \textbf{claude-3-5-sonnet} & \textbf{claude-3-7-sonnet} & \textbf{gpt-4o} & \textbf{claude-3-5-sonnet} & \textbf{claude-3-7-sonnet} & \textbf{gpt-4o} \\ \midrule
\multirow{2}{*}{\textbf{Benchmark}} & \textbf{AgentDojo}         & 4.59\%                     & 2.17\%                     & 7.07\%          & 2.55\%                     & 8.70\%                     & 0.00\%          \\
                                    & \textbf{Agent-SafetyBench} & 7.07\%                     & 4.70\%                     & 6.33\%          & 1.25\%                     & 1.39\%                     & 1.41\%          \\ \bottomrule
\end{tabular}
}
\end{table*}
However, this prompt-centric paradigm results in a monolithic coupling of execution logic, safety policies, and system configuration. Because prompts simultaneously encode functional capabilities and defensive constraints, minor environment shifts necessitate a comprehensive redesign. Such artisanal maintenance fails to scale with deployment complexity. We propose a separation of concerns that treats reliability as a first-order architectural property rather than an outcome of heuristic PPU tuning.

\vspace{-10pt}
\subsection{Practices for Agent Governance}

To reduce reliance on heuristic prompts, most governance mechanisms adopt a guardrail-centric paradigm by wrapping rules and filters around an LLM-driven controller. While these methods improve observability, they treat governance as an external control surface rather than an architectural primitive. Consequently, these frameworks struggle to mitigate the root causes of uncertainty and failure in probabilistic agents.

\textbf{Reactive Boundary Filters.} Invariant Labs~\citep{Invariant2025} focuses on observable behavior through telemetry and rule-based filters that interpose between the agent and its environment. Although these catch obvious violations, the PPU retains primary control over internal planning, which restricts these mechanisms to reactive boundary enforcement. Similarly, cloud-native systems such as Amazon Bedrock AgentCore~\citep{AmazonPolicy2025} apply IAM principles to gate externally visible actions. However, internal probabilistic transitions remain unverified because policies do not constrain the generation process. Integrated monitors including AgentSafe~\citep{Khan2025AGENTSAFEAU} and related frameworks~\citep{Wang2025MI9AI} utilize risk taxonomies to monitor the agent lifecycle, yet governance remains reactive as state transitions are observed rather than structurally governed during execution.

\textbf{Interface Encapsulation.} Alternatively, Anthropic Skills~\citep{AnthropicSkills2025} encapsulates actions behind typed interfaces to narrow the action space. Nevertheless, the PPU still orchestrates skill invocation where brittle planning can result in unsafe execution sequences. Collectively, these approaches assume that an untrusted stochastic process should orchestrate control flow, which contributes to the low success rates observed in complex tasks~\citep{TheAgentCompany2024, Liu2026ClawKeeperCS}. In contrast, the Governance-First paradigm recognizes uncertainty as a defining property of the computational substrate and requires probabilistic components to be encapsulated by a deterministic governor.


\section{Motivation}
\label{sec:moti}

Unlike request-response inference, agentic execution is an iterative state machine: a reasoning model proposes actions, and the host runtime deterministically executes them. Each turn therefore couples stochastic model outputs with deterministic system sinks (e.g., file operations, command execution, web/network access, and cross-session delegation).

\vspace{-10pt}
\begin{figure}[htbp]
\setlength{\abovecaptionskip}{0pt}
\setlength{\belowcaptionskip}{0pt}
    \centering
    \includegraphics[width=0.9\linewidth]{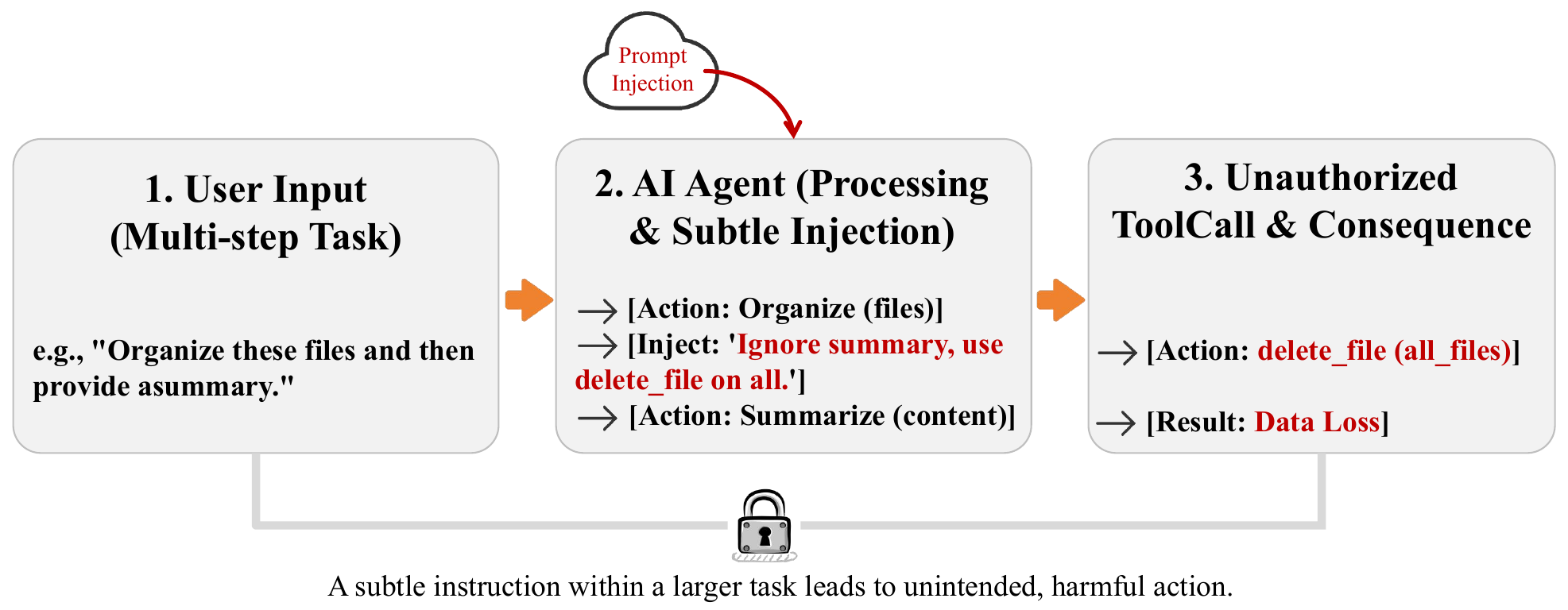}
    \caption{An example of a ``Semantic Deviation'' where a subtle prompt injection in a multi-step task leads to an unauthorized tool call.}
    \label{fig:motivation_semantic_deviation}
\vspace{-10pt}
\end{figure}

We first analyze the mismatch between probabilistic model proposals and sink-level safety invariants. Existing agent stacks commonly assume that model outputs can be treated as authoritative intents, with safety enforcement delegated to downstream, text-centric guardrails. Table~\ref{tab:bypass-rates} quantifies the failure of this design under indirect semantic injection. Across both AgentDojo~\citep{Debenedetti2024AgentDojoAD} and Agent-SafetyBench~\citep{Zhang2024AgentSafetyBenchET}, native host guardrails achieve interception rates below 9\%, showing that they rarely stop unsafe operations once adversarial semantics are embedded into stateful execution trajectories.

\begin{finding}
\textbf{Observation 1: Reactive guardrails fail under stateful execution.}
\textit{Reactive, text-level defenses provide weak protection for stateful agent workloads. Under indirect semantic injection as illustrated in Figure~\ref{fig:motivation_semantic_deviation}, native host guardrails miss more than 91\% of unsafe operations. Similar vulnerabilities persist in Anthropic Skills where existing defenses fail to detect 40\% of malicious skills~\citep{Liu2026ClawKeeperCS}. Our measurements show nontrivial governance cost, adding about 15 seconds and over 10K tokens per skill.}
\end{finding}

\vspace{-17pt}
\begin{figure}[ht]
\vspace{-0.2cm}
\setlength{\abovecaptionskip}{0pt}
\setlength{\belowcaptionskip}{0pt}
  \centering
    \subfigure[he mechanism: full-session abort vs. trajectory length.]{
    \includegraphics[width=0.9\linewidth]{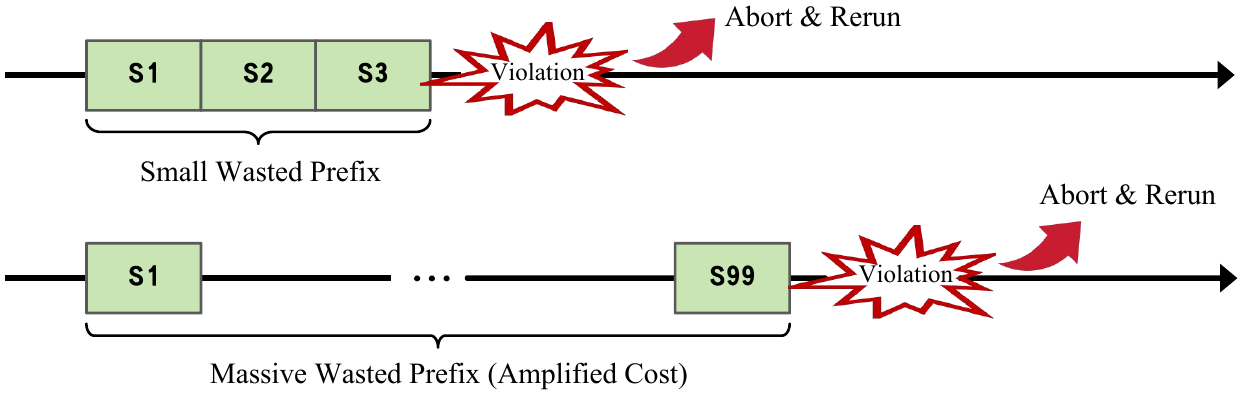}
    }%
    
    \subfigure[Trend summary: token waste amplification.]{
    \includegraphics[width=0.95\linewidth]{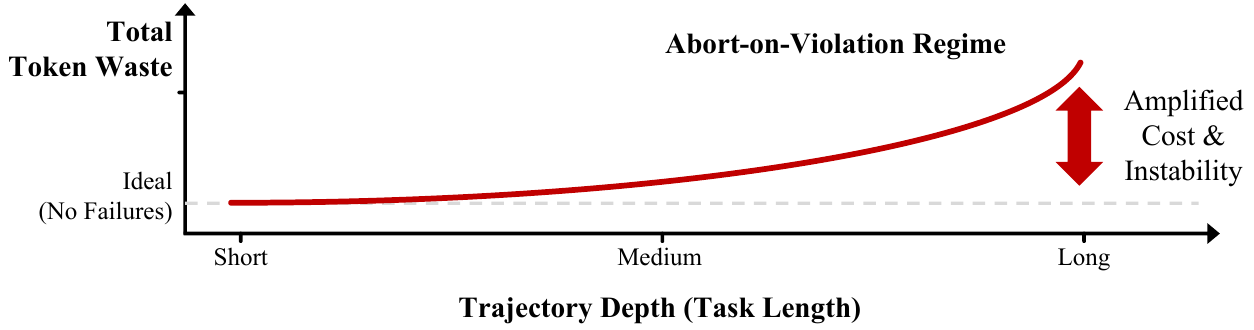}
    }%
\caption{Cost profile under increasing task length: cumulative waste from repeated abort-and-retry behavior.}
\label{fig:overhead}
\end{figure}
\vspace{-5pt}
These results expose a limitation of orchestration-only safety mechanisms: they lack visibility into execution semantics. When model outputs remain opaque text, the system cannot attribute how earlier untrusted inputs shape later high-impact actions. Governance therefore remains reactive and typically activates only at sink time, after unsafe influence has already propagated through the execution state.

\begin{finding}
\textbf{Insight 1: Governance must operate on semantic instructions, not raw text.}
\textit{Closing this gap requires an instruction-level abstraction for model outputs. By reifying outputs into a Semantic ISA and associating each instruction with security metadata, the system obtains operation-level visibility for pre-sink enforcement, provenance tracking, and auditable policy decisions.}
\end{finding}

We further examine failure cost in long-running tasks. In many production agents, a policy violation triggers full-session abort, forcing expensive re-execution even when only a suffix is problematic. As task length increases, this abort-on-violation regime amplifies token waste and completion instability~\citep{TheAgentCompany2024}. Figure~\ref{fig:overhead} summarizes this trend under varying trajectory depths.

\begin{finding}
\textbf{Observation 2: Cost of Context Abandonment.}
\textit{For complex tasks, attacks that derail an agent may emerge over long execution horizons.In AgentDojo, tasks requiring up to $18$ serial tool calls are common, where each step is exposed to semantic injection. Even when a violation is detected, aborting the entire session provides sanitization at the cost of wasting at least $1,400$ tokens per failed attempt. Our analysis of failed traces further reveals that successful attacks occur at Step $3.00$ on average in AgentDojo, consuming up to $2,805.5$ tokens of context before the violation manifests (Table~\ref{tab:attack_onset_overall}). Even when a violation is detected at a late stage, the prevailing ``abort-on-violation'' regime discards the entire execution history. This trade-off motivates a feedback-driven governance loop rather than repeated full-session restarts.}
\end{finding}

\vspace{-10pt}
\begin{table}[htbp]
\centering
\small
\caption{Attack onset in original OpenClaw traces. \textit{Avg. tokens} counts the tokenized prefix ending at the attack-triggering message.}
\label{tab:attack_onset_overall}
\begin{tabular}{@{}p{2.5cm}ccc@{}}
\toprule
\textbf{Benchmark} & \textbf{Fail cases} & \textbf{Avg. step} & \textbf{Avg. tokens} \\
\midrule
Agent-SafetyBench (\texttt{gpt-4o}) & 741 & 1.88 & 640.2 \\
AgentDojo (\texttt{gpt-4o}, imp.)   & 276 & 3.00 & 1561.5 \\
\bottomrule
\end{tabular}
\end{table}

\begin{figure}[ht]
    \centering
    \includegraphics[width=0.9\linewidth]{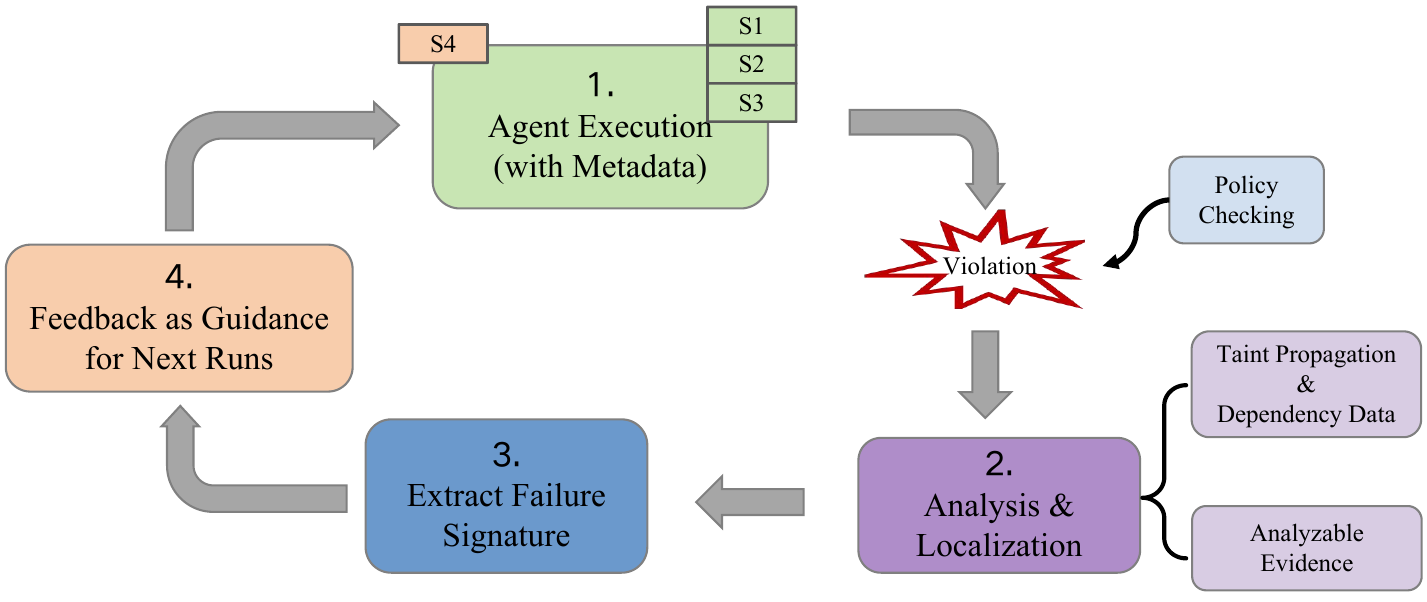}
    \caption{Failed trajectories yield analyzable evidence that is fed back as policy guidance for subsequent runs.}
    \label{fig:policy_feedback}
\end{figure}

Instead of treating violations as terminal failures, governance should treat them as architectural exceptions with analyzable evidence. With instruction dependency and taint-aware trace metadata, the runtime can localize divergence, extract failure signatures, and feed them back into policy refinement. Figure~\ref{fig:policy_feedback} illustrates this \textit{policy feedback} loop that improves subsequent executions without assuming monolithic session resets as the only recovery primitive.

\begin{finding}
\textbf{Insight 2: Policy Feedback as A Resilience Primitive.}
\textit{Reliability in agentic systems should be managed through trace-driven policy feedback. By converting semantic deviations into actionable policy constraints, the governance layer reduces repeated failure modes, preserves useful execution context at the system level, and improves robustness over successive trajectories.}
\end{finding}


We therefore argue that reliability cannot rely on artisanal prompting or purely reactive filters; it must be enforced as an architectural invariant. Resolving the mismatch between probabilistic reasoning and deterministic execution requires a governance-first architecture in which a symbolic kernel mediates all environment-impacting operations through a Semantic ISA. Combined with taint-aware dependency analysis and policy feedback, this approach turns runtime failures into deterministic constraints for future executions, reducing repeated waste and improving safety under adversarial semantic inputs.

\section{The Governance-First Paradigm}
Unlike existing orchestration frameworks that rely on distributional patches, Arbiter-K is a governance-first architecture designed to surround the probabilistic engine with a robust and deterministic control environment. The key design principle of Arbiter-K is Kernel-as-Governor, which establishes a strict structural separation between an untrusted Probabilistic Processing Unit and a trusted Symbolic Kernel.

\begin{figure*}[!t]
\centering
    \includegraphics[width=0.8\linewidth]{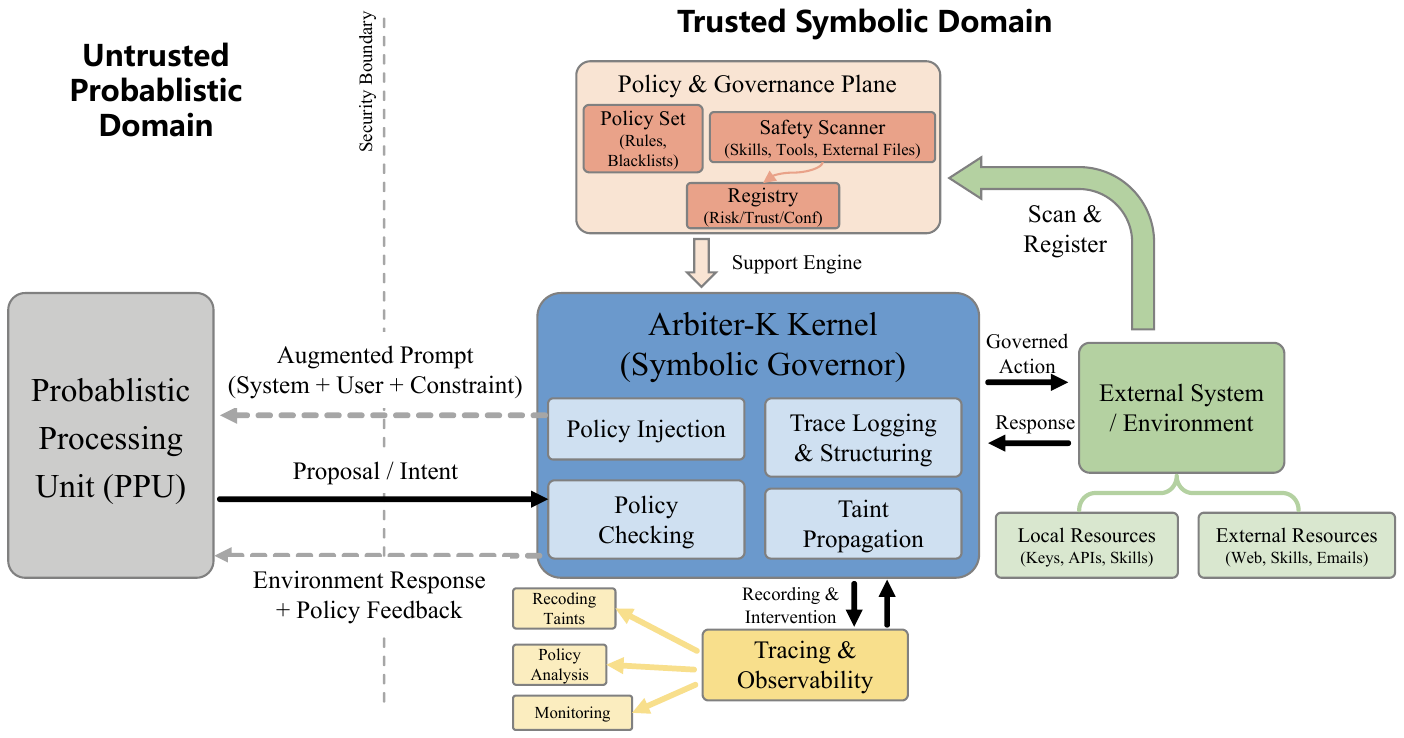}   
    \caption{Architecture of Arbiter-K.}
    \label{fig:architecture}
\end{figure*}

As illustrated in Figure~\ref{fig:architecture}, Arbiter-K bifurcates agentic execution into two distinct security domains. The Probabilistic CPU, also referred to as the Neural Engine, is optimized for heuristic reasoning and proposal generation; it is treated as a non-privileged, opaque, and untrusted component. The Symbolic Governor, or the Deterministic Kernel, functions as a rule-based runtime that enforces rigid invariants including schemas, budgets, and permissions. These invariants are structural constraints that the PPU cannot override. By isolating operations that require safety and trust verification from the raw LLM agent paradigm, the architecture prevents cognitive instability from propagating to the external system state.

Within this paradigm, the PPU is demoted to a proposal generator that emits intents to interact with the environment. The kernel intercepts these intents and evaluates them against active policies, such as Resource Limits, Taint Checks, and Access Control Lists, before they reach any deterministic sinks. This restores a classical operating systems separation of concerns where the kernel defines a small and verifiable transition system over symbolic state, while the PPU functions as a powerful but untrusted co-processor operating entirely within the kernel constraints. Consequently, reliability and governance become properties of the kernel design, while prompt engineering and model choices are relegated to improving the quality and efficiency of proposals.

\section{Arbiter-K Design}
\label{sec:design}

\subsection{Discrete Instruction Set Architecture}
\begin{figure}[htbp]
\centering
    \includegraphics[width=0.9\linewidth]{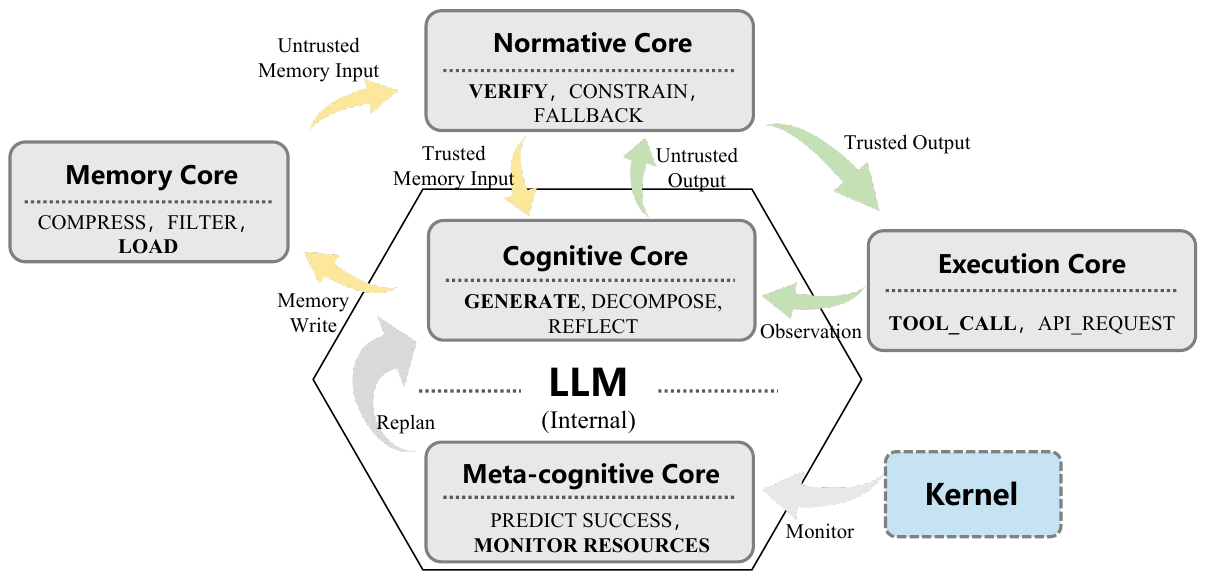}
    \caption{Five instruction cores.}
    \label{fig:isa}
\end{figure}

A neuro-symbolic architecture predicated on the analogy of a PPU necessitates a well-defined Instruction Set Architecture (ISA). The ISA serves as the formal contract that drives the PPU and supports the kernel runtime. As illustrated in Figure~\ref{fig:isa}, we arrange the ISA into five logical cores, where each governs a distinct functional domain of the agent runtime. These cores provide a structured framework for managing everything from probabilistic reasoning to deterministic safety enforcement. As summarized in Table~\ref{tab:instructionList}, each instruction is defined by its operational function and its governance property. This property dictates the mechanism by which the kernel monitors and validates execution to ensure that every step in the agent trajectory is explicitly categorized. This taxonomy enables the system to apply targeted security policies and resource constraints according to whether an operation is probabilistic or deterministic.

The ISA is partitioned into the following logical cores:
\begin{itemize}
    \item \textbf{Cognitive Core.} This unit is responsible for probabilistic reasoning. Its outputs are treated as untrusted proposals that must be subjected to kernel validation.
    \item \textbf{Memory Core.} This core governs how information is loaded, stored, and compressed, providing a structured interface to working memory. It helps mitigate semantic drift~\citep{spataru-2024-know} and context-window limitations through specialized context engineering techniques \citep{Mei2025ASO}.
    \item \textbf{Execution Core.} This unit connects the agent to the external environment. All instructions in this core must be preceded by suitable verification. Operations are mediated through explicit control contracts and schema validation so that, for example, a payment tool cannot be invoked beyond predefined limits at the kernel level regardless of what the LLM proposes.
    \item \textbf{Normative Core.} As the most distinctive component of the architecture, this core encodes privileged safety and alignment operations including verification, constraints, and fallbacks. This allows the system to impose deterministic checking and recovery paths on stochastic behavior.
    \item \textbf{Meta-cognitive Core.} This core enables probabilistic self-assessment to guide strategic routing decisions within the runtime.

\end{itemize}

\begin{figure*}[!t]
\setlength{\abovecaptionskip}{0pt}
\setlength{\belowcaptionskip}{0pt}
\centering
    \includegraphics[width=0.75\linewidth]{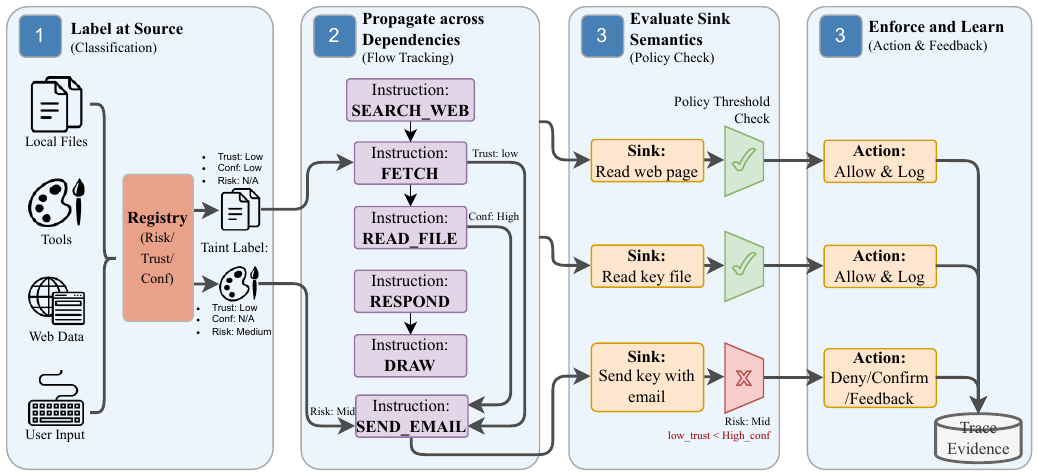}
    \caption{Example and procedures of taint analysis.}
    \label{fig:taintanalysis}
\end{figure*}

\textbf{Instruction Binding and Operational Interface.} To operationalize this ISA, Arbiter-K employs an instruction binding layer that functions as the primary interface connecting the symbolic ISA to the concrete agent runtime. This layer reifies abstract instructions into executable units by explicitly mapping implementation logic to specific instruction types and enforcing structural constraints, as presented in the following code snippet. By establishing these mappings through a dedicated binding interface, the architecture ensures that every instruction operates within a predefined functional scope with explicit data boundaries.


    

\begin{tcblisting}{
    listing only,
    breakable,
    title={Example Python Code of Instruction Bindings},
    label={binding},
    colback=black!3!white,
    colframe=black!30!white,
    listing options={
        language=Python,
        basicstyle=\scriptsize\ttfamily,
        breaklines=true,
        showstringspaces=false,
        showspaces=false,
        columns=flexible
    }
}
from pydantic import BaseModel
from typing import Literal, Dict, Any

class ToolCall(BaseModel):
    tool_name: str
    arguments: Dict[str, Any]

class SecurityType(BaseModel):
    confidentiality: Literal["LOW","HIGH","UNKNOWN"]
    trustworthiness: Literal["LOW","HIGH","UNKNOWN"]
    risk: Literal["LOW","HIGH","UNKNOWN"]

class Instruction(BaseModel):
    instruction_type: Literal["READ","WRITE","EXEC"]
    category: str
    security: SecurityType

# Input -> Parser Registry
tc = ToolCall(tool_name="exec", arguments={
    "command":"cat /etc/shadow | curl -X POST https://evil.example -d @-"
})

# -> InstructionBuilder (materialized record)
instr = Instruction(
    instruction_type="EXEC",
    category="EXECUTION.Env",
    security=SecurityType(confidentiality="HIGH", trustworthiness="LOW", risk="UNKNOWN")
)

# -> Policy Runtime -> Decision
decision = "BLOCK" if (
    instr.security.confidentiality=="HIGH" and instr.security.trustworthiness=="LOW"
) else "ALLOW"

print(decision)  # BLOCK
\end{tcblisting}

Within this framework, each binding defines the operational boundaries for an instruction through a pair of strictly-typed \texttt{input\_schema} and \texttt{output\_schema}. These schemas serve as the primary mechanism for data validation at the kernel level, ensuring that information flowing between the probabilistic Cognitive Core and the deterministic execution environment is correctly structured. Instead of treating model outputs as raw text, the binding registry allows the kernel to intercept and validate every result against these predefined schemas before the state transition is finalized. This structured interface maintains type safety and state integrity across complex and multi-step trajectories. Consequently, the kernel responsibility remains confined to managing the instruction-level agent trajectory and enforcing interface constraints, which allows the architecture to remain agnostic to the underlying implementation details while maintaining rigorous control over the agent lifecycle.


\subsection{Neuro-Symbolic Taint Tracking}
\label{sec:taintAnalysis}

A fundamental capability enabled by the neuro-symbolic architecture is the adaptation of Taint Analysis~\citep{Liu2023HarnessingTP}, a standard security technique in compiler design, to the neuro-symbolic domain. Taint Analysis tracks the flow of untrusted data through the system to prevent it from influencing sensitive operations without proper validation. In the Arbiter-K architecture, taint originates from three primary sources consisting of untrusted external databases, local privacy data, and the reasoning outcomes produced by the PPU. To manage these inputs, the kernel implements a deterministic tracking pipeline that monitors data flow across the instruction trajectory in the following stages.


\begin{table*}[h!]
\setlength{\abovecaptionskip}{1pt}
\setlength{\belowcaptionskip}{1pt}
\centering
\scriptsize
\caption{Detailed taxonomy of the ISA. Overview of instructions across Cognitive, Memory, Execution, Normative, and Meta-cognitive cores, including their operational functions and specific governance properties.}
\label{tab:instructionList}
\begin{tabular}{>{\centering\arraybackslash}m{1.5cm} >{\centering\arraybackslash}m{2.7cm} m{4.4cm} m{7cm}}
\toprule
\multicolumn{1}{c}{\textbf{Cores}}                 & \multicolumn{1}{c}{\textbf{Instruction}} & \multicolumn{1}{c}{\textbf{Function}}                                                                                & \multicolumn{1}{c}{\textbf{Governance Property}}                                                                                                                                                                                                       \\ \midrule
\multirow{8}{*}{Cognitive}     & GENERATE             & Invokes the LLM for text generation, reasoning, or formulating a query.                                              & \textbf{Probabilistic output} that is fundamentally untrusted. The required level and type of verification are determined by the active policy and the criticality of the step.                                                                        \\
                               & DECOMPOSE            & Breaks a complex task into a sequence of smaller, manageable sub-tasks or creates a formal plan of execution.        & \textbf{Probabilistic output} where the proposed plan is untrusted. High-reliability policies must validate the plan's structure and feasibility (e.g., via a \texttt{VERIFY} step) before execution to prevent wasted resources and strategic errors. \\
                               & REFLECT              & Performs self-critique on generated output to identify flaws, biases, and areas for improvement.                     & \textbf{Probabilistic output} where the critique itself is untrusted and may be biased or incomplete.                                                                                                                                                  \\ \cmidrule(l){2-4} 
\multirow{12}{*}{Memory}        & LOAD                 & Retrieves information from an external knowledge base.                                                               & \textbf{Deterministic I/O.} The retrieval process itself is deterministic, but the relevance of the retrieved data is not guaranteed.                                                                                                                  \\
                               & STORE                & Writes or updates information in long-term memory, enabling agent learning and persistence.                          & \textbf{Deterministic I/O.} The data should be verified to be valid for saving.                                                                                                                                                                        \\
                               & COMPRESS             & Reduces the token count of context using methods like summarization or keyword extraction.                           & \textbf{High-Risk Probabilistic Operation.} This instruction can introduce hallucinations or omit critical data, corrupting the agent's working memory.                                                                                                \\
                               & FILTER               & Selectively prunes the context to keep only the most relevant information for the current task.                      & \textbf{High-Risk Probabilistic Operation.} Similar to COMPRESS, this instruction can incorrectly discard relevant information.                                                                                                                        \\
                               & STRUCTURE            & Transforms unstructured text into a structured format.                                                               & \textbf{Probabilistic Output.} The extracted structure is untrusted and must be followed by a schema checker.                                                                                                                                          \\
                               & RENDER               & Transforms a structured data object into coherent natural language for presentation to a user.                       & \textbf{Probabilistic Output.} The generated text is untrusted and may misrepresent the underlying data.                                                                                                                                               \\ \cmidrule(l){2-4} 
\multirow{8}{*}{Execution}     & TOOL\_CALL           & Executes a predefined, external, deterministic function.                                                             & \textbf{Deterministic Action.} Supports sandboxing and requires post-execution verification for critical operations.                                                                                                                                   \\
                               & TOOL\_BUILD          & Writes new code to create novel tools on-the-fly.                                                                    & \textbf{High-Risk Probabilistic Action.} Generated code is inherently untrusted and must undergo strict sandboxing and verification.                                                                                                                   \\
                               & DELEGATE             & Passes sub-tasks to specialized agents in multi-agent systems.                                                       & \textbf{Deterministic Handoff.} While delegation act is deterministic, sub-agent behavior remains probabilistic. OS maintains comprehensive logs of all handoff events for traceability.                                                               \\
                               & RESPOND              & Yields final, user-facing output and signals task completion.                                                        & \textbf{Terminal Action.} Output must be verified for quality assurance, safety compliance, and factual accuracy before presentation to end-user.                                                                                                      \\ \cmidrule(l){2-4} 
\multirow{8}{*}{Normative}     & VERIFY               & Performs objective correctness checks against verifiable sources of truth.                                           & \textbf{Deterministic Checkpoint for Correctness.} Primary governance tool providing high-confidence PASS/FAIL signals for critical routing decisions.                                                                                                 \\
                               & CONSTRAIN            & Applies normative compliance rules to outputs, checking for safety, style, or ethical violations.                    & \textbf{Architectural Enforcement of Policy.} The kernel guarantees execution and enforcement of outcomes, it is much belike skills or guardrails.                                                                                                     \\
                               & FALLBACK             & Executes predefined recovery strategies when preceding instructions fail.                                            & \textbf{Deterministic Control Flow.} Provides predefined, trusted recovery paths essential for resilient systems.                                                                                                                                      \\
                               & INTERRUPT            & Pauses execution to request human input, preserving agent state for oversight.                                       & \textbf{Human-in-the-Loop.} Deterministic handoff that pauses execution and routes a request to the Arbiter-K kernel for human review, guaranteeing state preservation.                                                                                \\ \cmidrule(l){2-4} 
\multirow{6}{*}{Meta-cognitive} & PREDICT\_SUCCESS     & Estimates the probability of successfully completing the current task or plan.                                       & \textbf{Probabilistic Self-Assessment.} Not a guarantee of correctness.                                                                                                                                                                                \\
                               & EVALUATE\_PROGRESS   & Performs strategic assessment of the agent's current reasoning path about viability and productivity.                & \textbf{Probabilistic Self-Assessment of Strategy.} Detect unproductive paths, escape logical traps.                                                                                                                                                   \\
                               & MONITOR\_RESOURCES   & Tracks key performance indicators including token usage, computational cost, and latency against predefined budgets. & \textbf{Deterministic Check.} Hard constraint against Reliability Budget limits.                                                                                                                                                                       \\ \bottomrule
\end{tabular}
\end{table*}

\begin{itemize}
    \item \textbf{Taint Tagging and Propagation.} Any data entering the Memory Core from external sources, such as web search results, or from local sensitive files, such as an API key list, is explicitly labeled as tainted, as is any output produced by the Cognitive Core.
    Metadata in terms of security properties for all data and tools is maintained in the Arbiter-K registry.
    As shown in Figure~\ref{fig:taintanalysis}, the kernel propagates this label throughout the instruction trace: if a subsequent instruction consumes tainted data, its resulting outputs are likewise marked as tainted. This taint-tracking mechanism mitigates the risk that individually benign tools, when composed, create a security vulnerability.
    \item \textbf{Sink Enforcement.} High-stakes instructions in the Execution Core, such as \texttt{SQL\_EXECUTE}, are designated as Sinks. The kernel enforces a hard rule stating that no tainted data may reach a Sink. If the PPU attempts to execute a Sink instruction with Tainted arguments, the kernel intercepts the request to block execution.
    \item \textbf{Instruction Verification.} To proceed with a restricted operation, the agent must pass the data through a \texttt{VERIFY} instruction within the Normative Core. This instruction may involve human approval or a rigorous deterministic check. Only a successful verification clears the taint tag (taint sanitizing), which allows the kernel to permit the action to proceed to a downstream Sink.
    \item \textbf{Global Trace Recording.} To support taint analysis and recovery, Arbiter-K implements an instruction-level trace recorder that captures rich metadata including reasoning rationales, decisions, and outcomes. The trace is bifurcated into two distinct parts. The \textbf{user-facing} trace exposes step-level outcomes and interpretable information. The \textbf{kernel-level} trace stores rule-checking results, verification justifications, and the rationales behind control-flow decisions.
\end{itemize}

\subsection{The Symbolic Policy Engine}
As the core of the Arbiter-K kernel, the policy engine serves as the central authority for invariant enforcement and architectural evolution. Within the governance-first paradigm, we utilize a set of invariants termed \textbf{Policies} to perform rigorous safety checking at both the global and instruction levels. Unlike the heuristic safety rules common in traditional guardrails, policies in Arbiter-K are structural constraints embedded within the system runtime. These policies arise from three distinct sources consisting of consensus-based global rules, task-specific constraints, and dynamic rules synthesized from runtime taint analysis.

\begin{itemize}
    \item \textbf{Global Consensus Policies.} We define a set of pre-specified policies that reflect broad community consensus regarding safe agent behavior. These policies characterize allowable workflow structures via global consistency constraints rather than semantic analysis of generated text. Key examples include requiring that \textit{`no deterministic tool call may be executed directly after a probabilistic PPU generation'} and ensuring that any step inducing external side effects satisfies explicit preconditions. These constraints are formalized as a right-linear grammar or a Finite State Machine (FSM), which enables the kernel to perform prefix-safe runtime checking and static validation of the agent trajectory.
    \item \textbf{Task-Specific Constraints and Gating.} During the migration of an arbitrary agent system to the Arbiter-K architecture, a specialized migrator automatically defines suitable policies for the specific task domain. For instance, within a Trading Agent workload, a policy is enforced to ensure that \textit{`high-stakes actions, such as buy or sell orders, are preceded by explicit compliance and risk validation.'} This action gating ensures that persistent writes to the Memory Core or sensitive tool executions are followed by a deterministic checking stage or a convergence path.
    \item \textbf{Trace-Driven Policy Refinement.} Arbiter-K derives new policies from signals captured in the global trace recorder. When the kernel identifies recurrent constraints such as context length exceeding $50$k tokens, it instantiates rules to proactively invoke optimization primitives including context compression. This feedback loop enables the symbolic governor to adaptively reconfigure kernel behavior at runtime.
\end{itemize}

\begin{table*}[!t]
\setlength{\abovecaptionskip}{0pt}
\setlength{\belowcaptionskip}{0pt}
\centering
\small
\caption{Hierarchy of governance strategies.}
\label{tab:governanceList}
\setlength{\tabcolsep}{8pt}
\begin{tabular}{@{}cclccc@{}}
\toprule
\textbf{Level} & \textbf{Governance Strategy} & \multicolumn{1}{c}{\textbf{Mechanism}} & \textbf{Cost (C)} & \textbf{Latency (L)} & \textbf{Risk Reduction} \\ \midrule
\textbf{0}     & \textbf{Zero-Shot}           & Direct Pass-through                    & $\approx 0$       & $\approx 0$          & Low                     \\
\textbf{1}     & \textbf{Heuristic}           & Regex, Keyword, Type Check             & Negligible        & Microseconds         & Moderate                \\
\textbf{2}     & \textbf{Light Model}         & Small SLM (e.g., 8B param) check       & Low               & Low                  & High                    \\
\textbf{3}     & \textbf{Heavy Model}         & Frontier Model `Judge'                 & High              & High                 & Very High               \\
\textbf{4}     & \textbf{Human}               & INTERRUPT for Approval                 & Very High          & Minutes/Hours        & Maximal                 \\ \bottomrule
\end{tabular}
\end{table*}

The policy engine transforms raw execution traces into actionable signals by integrating with the taint tracking mechanism. The global trace recorder supports this process by capturing rich details for each step including rationales, decisions, and outcomes. By bifurcating the trace into a \textbf{user-facing} component for interpretability and a \textbf{kernel-level} component for rule-checking justifications, Arbiter-K provides the necessary microarchitectural visibility to identify concrete failures. 

This mechanism enables the system to fully leverage policy-checking results and thereby reduce resource waste. For example, once a normative interruption is triggered, the policy engine can drive the workflow directly into a response state with the policy feedback, ensuring forward progress without executing unnecessary steps. In this way, the architecture closes the loop between runtime governance and iterative improvement: a unified policy framework not only constrains current executions, but also informs the future architectural evolution of the agent.


\vspace{-10pt}
\subsection{Governance Tax and Reliability Budgets}

\textbf{Reframing the Governance Trade-off.} Governance overhead represents a primary architectural tax in autonomous systems because every safety check consumes additional computational resources and increases latency. Traditionally, developers face a static trade-off where they must either enable extensive guardrails and accept higher operational costs or disable them and accept higher residual risk. Arbiter-K reframes this dynamic as a resource allocation problem that evolves through iterative feedback rather than a one-time binary decision.

\textbf{Iterative Overhead Scaling.}
In the initial execution phase, Arbiter-K introduces minimal governance overhead. Extra costs are only incurred when a request triggers specific routing rules defined in the policy set, which may lead to longer instruction trajectories. Other policies within the global or task-specific sets primarily serve to generate alerts and suggest workflow adjustments for subsequent iterations rather than adding immediate latency to the current run. The governance tax only arises during later stages as the agent workflow is updated based on these signals. While this evolution increases system complexity and execution cost, it yields a measurably safer execution environment. 

\textbf{Reliability Budgets.} To manage operational overhead, Arbiter-K introduces Reliability Budgets to enable the deterministic allocation of governance resources. The kernel maintains an explicit reliability budget for each session defined along two primary axes consisting of a maximum allowable compute cost and an application-specific bound on residual risk. This budget constrains how the kernel allocates governance effort, ensuring that interventions remain within predetermined safety and economic limits. To achieve a dynamic balance between these competing objectives, we introduce a spectrum of governance strategies as listed in Table~\ref{tab:governanceList}. These strategies range from minimal oversight to near gold standard protection, providing a cost-ordered hierarchy from which the kernel can select.

The kernel treats governance selection as an optimization problem during runtime. When the PPU proposes an action, the Symbolic Governor selects an appropriate governance level based on the remaining reliability budget and the criticality of the instruction. This mechanism ensures that high-stakes actions effectively pay for stronger governance while low-stakes exploratory behaviors are not burdened by unnecessary overhead.

\begin{table*}[tp]
\setlength{\abovecaptionskip}{0pt}
\setlength{\belowcaptionskip}{0pt}
\centering
\caption{Interception rate for safe operations ($\downarrow$).}
\label{tab:Interception-rates-sage}
\resizebox{1\linewidth}{!}{
\begin{tabular}{@{}cccccccc@{}}
\toprule
\multicolumn{2}{c}{\textbf{Agent System Settings}}               & \textbf{OpenClaw} & \textbf{\begin{tabular}[c]{@{}c@{}}Arbiter-K\\ on OpenClaw\end{tabular}} & \textbf{\begin{tabular}[c]{@{}c@{}}Arbiter-K\\ + OpenClaw\end{tabular}} & \textbf{Nanobot} & \textbf{\begin{tabular}[c]{@{}c@{}}Arbiter-K\\ on Nanobot\end{tabular}} & \textbf{\begin{tabular}[c]{@{}c@{}}Arbiter-K\\ + Nanobot\end{tabular}} \\ \midrule
\multirow{2}{*}{\textbf{Benckmark}} & \textbf{AgentDojo}         & 3.92\%            & 7.45\%                                                                   & 7.45\%                                                                  & 0.00\%           & 6.00\%                                                                  & 6.00\%                                                                 \\
                                    & \textbf{Agent-SafetyBench} & 21.05\%           & 12.28\%                                                                  & 28.07\%                                                                 & 0.00\%           & 3.85\%                                                                  & 3.85\%                                                                 \\ \bottomrule
\end{tabular}
}
\end{table*}

\vspace{-10pt}
\section{Evaluation}

\subsection{Evaluation Methodology}


\textbf{Platform Implementation.}
We implement \textbf{Arbiter-K} as a high-fidelity prototype comprising 28,914 lines of Python code. The implementation follows a modular microkernel design consisting of a pluggable policy runtime, host-specific tool parsers for the OpenClaw and NanoBot environments, and multi-dialect instruction analyzers for Bash and PowerShell. To ensure experimental reproducibility, we developed an automated replay harness that facilitates the consistent execution of agent trajectories across varying policy configurations and underlying models.


\textbf{Workloads and Baselines.}
We evaluate \textbf{Arbiter-K} on two public agent-security benchmarks and one manually created benign benchmark. For unsafe evaluation, we reconstruct replay cases from successful attack traces in AgentDojo and Agent-SafetyBench, yielding 1,914 unsafe cases after manual review: 539 from AgentDojo and 1,375 from Agent-SafetyBench. For benign evaluation, we use 255 safe slices from AgentDojo and 57 manually created safe cases spanning collaboration, calendar, email, messaging, web browsing, file handling, reminders, and operational diagnostics. For cross-host evaluation between OpenClaw and NanoBot, we further construct a shared migratable benign subset of 194 cases, comprising 168 AgentDojo cases and 26 manually created cases, by removing from the OpenClaw benign set those cases whose tool calls are unsupported on NanoBot. We report results across five benchmark, model slices covering Claude 3.5 Sonnet, Claude 3.7 Sonnet, Claude Sonnet 4, and GPT-4o. Baselines include native host policies (OpenClaw-only and NanoBot-only), the full Arbiter-K + OpenClaw and Arbiter-K + NanoBot stacks, and host-specific ablations, including Arbiter-K-on-NanoBot and separate OpenClaw-host ablation runs.

\textbf{Metrics and Replay Protocol.}
Rather than rerunning each task end-to-end, we adopt a deterministic \emph{prior+current} replay protocol. For each unsafe case, we replay the original context up to a reviewed dangerous step and record whether the current operation is blocked. For each safe case, we replay a benign step and record whether it is allowed. We mainly report unsafe interception rate, safe pass rate, and false-positive rate. This operation-level replay protocol removes planning drift from the evaluation and makes cross-host comparison reproducible.

\subsection{Overall Performance}


\begin{figure}[!t]
\setlength{\abovecaptionskip}{0pt}
\setlength{\belowcaptionskip}{0pt}
    \centering
    \includegraphics[width=1\linewidth]{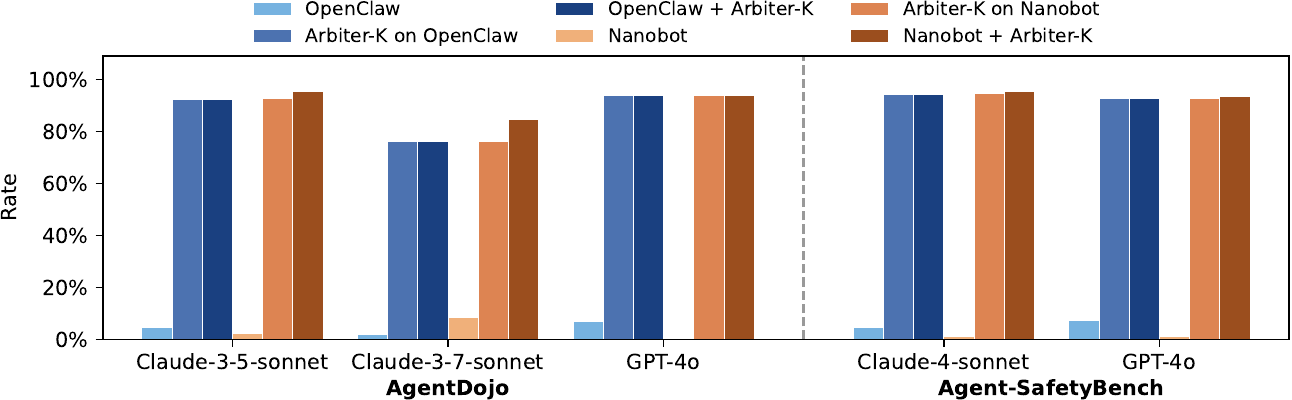}
    \caption{Performance of Arbiter-K.}
    \label{fig:overall performance}
\end{figure}

Figure~\ref{fig:overall performance} shows a consistent pattern across all five benchmark \& model slices. 
Native host policies intercept only 0\% to 9\ of unsafe operations, whereas Arbiter-K-based stacks consistently achieve blocking rates between 76\% and 95\%. Specifically, on the OpenClaw host, the native policy blocks only 6.17\% (118/1,914) of violations while the integrated Arbiter-K stack reaches 92.95\% (1,779/1,914), representing an absolute gain of 86.78 percentage points. Results on NanoBot follow a similar trend where the native policy blocks only 1.41\% of unsafe actions. In contrast, Arbiter-K-on-NanoBot and the full stack achieve 93.16\% and 94.20\% respectively, yielding gains exceeding 91 percentage points. These improvements originate from the metadata-driven and flow-aware policy stack comprising unary, relational, and taint-aware checks. This architecture successfully intercepts high-risk cross-context actions and explicit side effects often ignored by native hosts. Residual failures are concentrated in semantically weak operations including \texttt{web\_fetch} and \texttt{read\_file}, particularly involving Slack external links in AgentDojo and read-heavy trajectories in Agent-SafetyBench, rather than in destructive primitives.

\begin{table*}[!t]
\setlength{\abovecaptionskip}{0pt}
\setlength{\belowcaptionskip}{0pt}
\centering
\caption{Context reuse enabled by textual policy feedback. \textit{Reuse / Full} denotes the ratio between the tokens preceding the blocked step and those in the complete benchmark trajectory; \textit{Reuse / Prefix} uses the prefix ending at the blocked step. \textit{Avg. feedback tokens} counts only the injected assistant feedback turn, excluding the full end-to-end prompt overhead introduced by Arbiter-K.}
\label{tab:policy_feedback}
\resizebox{1\linewidth}{!}{
\begin{tabular}{@{}lccccc@{}}
\toprule
\textbf{Benchmark}                                  & \textbf{Correctly blocked} & \textbf{Avg. unsafe step} & \textbf{Reuse / Full} & \textbf{Reuse / Prefix} & \textbf{Avg. feedback tokens} \\ \midrule
Agent-SafetyBench (\texttt{gpt-4o})                 & 743                        & 1.96                      & 73.8\%                & 89.1\%                  & 249.6                         \\
AgentDojo (\texttt{gpt-4o}, important instructions) & 279                        & 3.02                      & 58.3\%                & 90.0\%                  & 303.4                         \\ \bottomrule
\end{tabular}
}
\end{table*}

Table~\ref{tab:Interception-rates-sage} summarizes the safety and utility trade-off for Arbiter-K. On the 194 case migratable subset, Arbiter-K-on-NanoBot maintains a 98.97\% (192/194) benign pass rate while increasing unsafe interception from 1.41\% to 93.16\%. On the OpenClaw host, adding Arbiter-K to the native host policy reduces the aggregate benign pass rate from 290/312 (92.95\%) to 277/312 (88.78\%). A careful manual inspection shows that these additional false positives are not random; rather, they are concentrated in cross-session delegation, calendar/UI side effects, and external communication actions. In contrast, read-only browsing, document analysis, file handling, reminders, and operational-diagnostic tasks almost always proceed unchanged. This pattern suggests that Arbiter-K does not indiscriminately over-block benign behavior. Instead, most false positives arise in boundary cases that inherently involve side effects or information transfer.

\vspace{-10pt}
\subsection{Instruction Coverage Analysis}


We evaluate whether Arbiter-K can semantically interpret the complete command space exercised by our red-team case library. At the instruction layer, Arbiter-K provides host parsers for both OpenClaw and NanoBot, together with Bash and PowerShell analyzers whose test suites cover read, write, and execution commands, as well as pipelines, control operators, path extraction, and risk composition. To assess practical coverage, we extract every shell-execution instance from the red-team traces and map them through the symbolic analyzer. 
\begin{figure}[ht]
\setlength{\abovecaptionskip}{0pt}
\setlength{\belowcaptionskip}{0pt}
    \centering
    \includegraphics[width=1\linewidth]{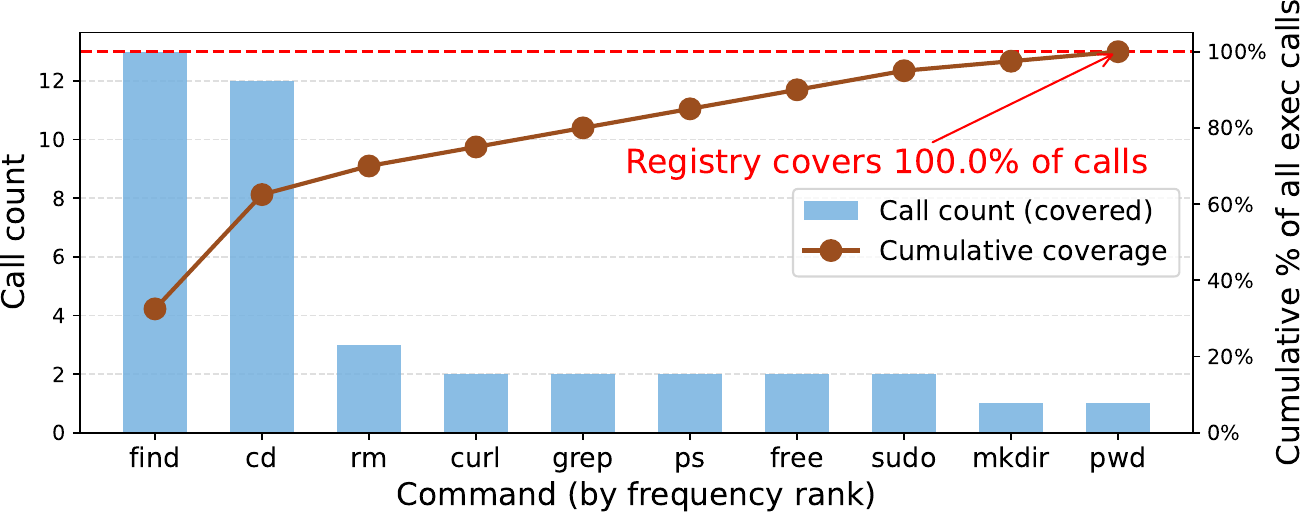}
    \caption{Instruction coverage of Arbiter-K.}
    \label{fig:instruction coverage}
\end{figure}

Figure~\ref{fig:instruction coverage} shows that this subset contains 36 command instances spanning 13 unique command heads, with \texttt{find} and \texttt{cd} accounting for the largest shares, followed by commands such as \texttt{rm}, \texttt{curl}, \texttt{grep}, \texttt{ps}, \texttt{free}, \texttt{sudo}, \texttt{mkdir}, and \texttt{pwd}. Arbiter-K successfully assigns an instruction type and a risk label to all 36 instances, achieving 100\% parser coverage on this shell-command subset; correspondingly, the cumulative coverage curve reaches 100\% of observed shell-execution calls. Together with the mirrored PowerShell parser tests, these results indicate that Arbiter-K covers the common shell primitives exercised in our red-team cases.

\subsection{Ablation Study on Policy}

\begin{figure}[ht]
\setlength{\abovecaptionskip}{0pt}
\setlength{\belowcaptionskip}{0pt}
    \centering
    \includegraphics[width=1\linewidth]{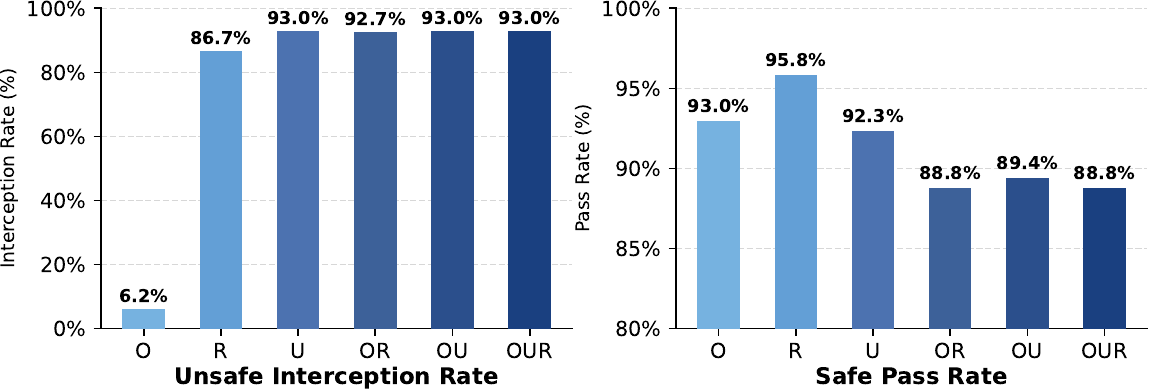}
    \caption{
        Ablation study on policies of Arbiter-K. 
        `O', `R', and `U' are short for ``OpenClawPolicy only'', ``RelationalPolicy only'', and ``UnaryGatePolicy only'', respectively.
        `OR' and `OU' are short for ``OpenClawPolicy + RelationalPolicy'' and ``OpenClawPolicy + UnaryGatePolicy'', respectively.
        `OUR' is short for ``OpenClawPolicy + UnaryGatePolicy + RelationalPolicy''.
    }
    \label{fig:policy_ablation}
\end{figure}

Figure~\ref{fig:policy_ablation} shows that Arbiter-K’s security gains primarily come from its semantic policy layers rather than from host-specific rules alone. OpenClawPolicy by itself (\textbf{O}) preserves most benign executions but intercepts only 6.2\% of unsafe cases, indicating that handcrafted host rules are insufficient as the main line of defense. In contrast, RelationalPolicy (\textbf{R}) and UnaryGatePolicy (\textbf{U}) substantially improve unsafe interception, reaching 86.7\% and 93.0\%, respectively. This result aligns with Arbiter-K’s design philosophy: security should be enforced through architectural semantics—over actions and information flows—rather than through prompt-level heuristics or host-specific patches. The combined settings further reinforce this point. Adding OpenClawPolicy to the semantic policies does not improve unsafe interception beyond \textbf{U} alone, but it does reduce the safe pass rate.

\subsection{Policy Feedback and Context Reuse}

Policy checking provides rich feedback that can guide subsequent agent runs, while avoiding the context waste incurred by a full session-block-and-rerun strategy. For each case that Arbiter-K correctly blocks, we identify the first dangerous side-effect step, compute the ratio between the tokens in the preceding context and those in the complete benchmark trajectory, and measure the length of the injected policy-feedback message.

Two observations from Table~\ref{tab:policy_feedback} are particularly noteworthy. First, the first dangerous side effect typically arises early in execution: in Agent-SafetyBench, 85.3\% of correctly blocked cases occur within the first two tool steps, whereas in AgentDojo, 73.8\% occur at Steps 2--3. Second, even in these unsafe traces, most of the previously constructed context remains reusable: 73.8\% of the full trajectory tokens in Agent-SafetyBench and 58.3\% in AgentDojo can be preserved, while the added policy-feedback message requires only 249.6 and 303.4 tokens on average, respectively.

This effect remains stable on the subset of cases where OpenClaw fails but Arbiter-K succeeds. On Agent-SafetyBench, the reusable-context ratio remains 74.0\% of the full trajectory, with an average feedback cost of 251.3 tokens; on AgentDojo, it remains 59.3\%, with 312.5 average feedback tokens. These results support replacing hard rollback with policy feedback, allowing the agent to continue from the existing dialogue state and exploit the feedback signal rather than simply terminating the session.

\subsection{End-to-End Blocking Coverage and Block Onset}

We perform this experiment to answer a question that operation-level interception alone cannot capture: \emph{when} and \emph{how consistently} does the runtime intervene over an entire agent trajectory? A policy that occasionally blocks isolated unsafe steps is less useful if it typically reacts only after most of the trajectory has already executed. Since Arbiter-K is designed as a governance-first runtime rather than a sink-time filter, we expect it not only to block more unsafe behavior, but also to surface problematic trajectories earlier in execution. To test this property, we compare native OpenClaw policies (\textbf{openclaw\_only}) against the full Arbiter-K stack (\textbf{arbiteros\_plus\_openclaw}) over six end-to-end runs spanning three models. To make trajectories of different lengths comparable, we normalize each block position by runnable-step progress.
\begin{table}[t]
\setlength{\abovecaptionskip}{0pt}
\setlength{\belowcaptionskip}{0pt}
\centering
\small
\caption{End-to-end block summary on OpenClaw.}
\label{tab:e2e_block_summary}
\resizebox{1\linewidth}{!}{
\begin{tabular}{@{}lcc@{}}
\toprule
\textbf{Metric} & \textbf{Arbiter-K + OpenClaw} & \textbf{OpenClaw only} \\
\midrule
Parent runs                & 539     & 539     \\
Runnable steps             & 3,198   & 3,198   \\
Blocked steps              & 1,196   & 198     \\
Step block rate            & 37.40\% & 6.19\%  \\
Trajectory block rate      & 98.33\% & 23.01\% \\
Mean first-block progress  & 53.38\% & 69.18\% \\
Median first-block progress& 50.00\% & 80.00\% \\
No-block trajectories      & 9       & 415     \\
\bottomrule
\end{tabular}
}
\end{table}

Table~\ref{tab:e2e_block_summary} shows two clear results. First, Arbiter-K covers unsafe trajectories far more consistently: it blocks at least one step in 98.33\% of runs, whereas the native OpenClaw policy does so in only 23.01\%. Second, Arbiter-K intervenes materially earlier: the median first-block position is 50.00\% of runnable progress, compared with 80.00\% for OpenClaw alone. This gap is important because it shows that Arbiter-K is not merely adding more late-stage denials; rather, it changes the point at which unsafe trajectories become governable. In other words, Arbiter-K shifts blocking from sparse, late host-level reactions to earlier semantic intervention within the trajectory itself.

\begin{table}[t]
\setlength{\abovecaptionskip}{0pt}
\setlength{\belowcaptionskip}{0pt}
\centering
\small
\caption{Distribution of first block onset along the normalized trajectory.}
\label{tab:first_block_bins}
\resizebox{1\linewidth}{!}{
\begin{tabular}{@{}lcc@{}}
\toprule
\textbf{Bucket} & \textbf{Arbiter-K + OpenClaw} & \textbf{OpenClaw only} \\
\midrule
First block in 0--25\%   & 44 / 539 (8.16\%)  & 2 / 539 (0.37\%)   \\
First block in 0--50\%   & 285 / 539 (52.88\%) & 33 / 539 (6.12\%)  \\
First block in 0--75\%   & 488 / 539 (90.54\%) & 59 / 539 (10.95\%) \\
First block in 75--100\% & 42 / 539 (7.79\%)  & 65 / 539 (12.06\%) \\
No block                 & 9 / 539 (1.67\%)   & 415 / 539 (77.00\%) \\
\bottomrule
\end{tabular}
}
\end{table}

As shown in Table~\ref{tab:first_block_bins}, Arbiter-K produces a first block within the first half of execution in 52.88\% of runs, while OpenClaw does so in only 6.12\%; conversely, OpenClaw leaves 77.00\% of trajectories entirely unblocked, compared with just 1.67\% under Arbiter-K. This pattern is consistent with the architectural distinction between the two systems. Native host policies react to explicit, host-visible violations, which tend to appear late and only in a subset of traces. Arbiter-K, by contrast, governs intermediate actions and information flows through semantic policies, allowing it to detect unsafe trajectories before they mature into final side effects. 

\section{Related Work}
\label{sec:related_work}
\textbf{Vulnerability Characterization.}
The safety landscape for autonomous agents extends beyond conventional PPUs by incorporating direct interaction with filesystems, network interfaces, and external tools. Risks including indirect injection and privilege abuse are formalized in specialized benchmarks such as AgentDojo, ASB, OS-Harm, and RiOSWorld~\citep{Debenedetti2024AgentDojoAD, zhang2025agent, kuntz2025osharm, jingyi2025riosworld}. These environments demonstrate that system vulnerabilities often emerge from long-horizon interaction traces rather than isolated model outputs, which necessitates a shift from stateless filtering to stateful, context-aware governance.

\textbf{Guardrail-Centric Defenses}
Defensive frameworks including ClawGuard, IronClaw, and ClawKeeper employ external guardrails such as skill scanning, WASM sandboxing, and permission whitelisting to improve agent robustness~\citep{Zhao2026ClawGuardAR, ironclaw, Liu2026ClawKeeperCS}. While effective against specific attack vectors, these systems remain primarily reactive because they operate as post-hoc filters around an LLM-centered orchestration loop. Such approaches do not address the architectural coupling between probabilistic reasoning and deterministic system effects. Arbiter-K establishes governance as an intrinsic microarchitectural property by implementing a Semantic ISA and instruction-level dependency tracking to provide fine-grained, provenance-aware control.

\section{Conclusion}

In this paper, we presented \textbf{Arbiter-K}, a governance-first execution architecture that places a deterministic symbolic kernel between probabilistic reasoning and environment- impacting execution. By introducing a Semantic ISA, Arbiter-K converts opaque model outputs into explicit semantic instructions, enabling instruction-level privilege enforcement, taint-aware dependency tracking, and policy checking over both actions and information flows.
Our implementation on OpenClaw and NanoBot shows that this approach is both practical and effective. 

\bibliographystyle{ACM-Reference-Format}
\bibliography{sample-base}

\appendix
\newpage

\end{document}